\documentclass[aps,prl,preprint,groupedaddress,floatfix]{revtex4}
\usepackage{graphicx}
\begin{document}

\title{Half-life and spin of $^{60}$Mn$^{g}$}

\author{S.N.~Liddick$^{1,2}$, P.F.~Mantica$^{1,2}$, 
B.A.~Brown$^{1,3}$, M.P.~Carpenter$^{4}$,
A.D.~Davies$^{1,3}$, M.~Horoi$^{5}$,  
R.V.F.~Janssens$^{4}$, A.C.~Morton$^{1}$, W.F.~Mueller$^{1}$,  
J.~Pavan$^{6}$, H.~Schatz$^{1,3}$, A.~Stolz$^{1}$, 
S.L.~Tabor$^{6}$, B.E.~Tomlin$^{1,2}$, M.~Wiedeking$^{6}$}

\affiliation{$^{(1)}$ 
National Superconducting Cyclotron
Laboratory, Michigan State University,
East Lansing, Michigan 48824}
\affiliation{$^{(2)}$
Department of Chemistry, Michigan State University,
East Lansing, Michigan 48824}
\affiliation{$^{(3)}$
Department of Physics and Astronomy,
Michigan State University, East Lansing, Michigan
48824}
\affiliation{$^{(4)}$ 
Physics Division,
Argonne National Laboratory, Argonne, Illinois 60439}
\affiliation{$^{(5)}$
Department of Physics, Central Michigan University,
Mount Pleasant, Michigan 48859}
\affiliation{$^{(6)}$
Department of Physics, Florida State University,
Tallahassee, Florida 32306}

\date{\today}

\begin{abstract}
A value of $0.28\pm 0.02$~s has been deduced for 
the half-life of the ground state of $^{60}$Mn, in 
sharp contrast to the previously adopted value of $51 \pm 6$~s.
Access to the low-spin $^{60}$Mn ground state was 
accomplished via $\beta$ decay of the $0^+$ $^{60}$Cr 
parent nuclide.  New, low-energy states in $^{60}$Mn
have been identified from $\beta$-delayed
$\gamma$-ray spectroscopy.  The new, shorter half-life of
$^{60}$Mn$^{g}$ is not suggestive of isospin forbidden $\beta$ 
decay, and new spin and parity assignments of $1^+$ and 
$4^+$ have been adopted for the ground and isomeric $\beta$-decaying
states, respectively, of $^{60}$Mn.
\end{abstract}
\pacs{}
\maketitle

\section{Introduction}
The nuclide $^{60}$Mn has two known $\beta$-decaying states.  
$\beta$ decay of a proposed $3^+$ level was initially established
by Norman {\it et al.} \cite{nor1978}, and a half-life 
of $1.79\pm 10$~s was reported together with several
delayed $\gamma$-ray transitions depopulating excited
states in the $^{60}$Fe daughter nuclide.  Subsequently,
Runte {\it et al.} \cite{run1985} deduced that the 
$3^+$ level was isomeric, decaying with a internal transition
(IT) to $\beta$ ratio
of $0.13 \pm 0.01$.  A 272-keV $\gamma$ ray, with a 
decay half-life of 1.8~s, was designated as the IT, and 
$M3$ multipolarity was assigned based on Weisskopf estimates. 
Additional $\beta$-delayed $\gamma$ rays were identified
by Runte {\it et al.}, including a 1150-keV transition
known to depopulate the first excited $0^+$ state in 
$^{60}$Fe.  The first measurement of the half-life of the 
presumed $0^+$ ground state of $^{60}$Mn was completed 
by Bosch {\it et al.} \cite{bos1988}.  
A half-life of $51 \pm 6$~s was deduced from the multiscaled 
$\beta$ singles counting rate.  
No evidence was found in their work for delayed 
$\gamma$ rays following $\beta$ decay 
of the $^{60}$Mn ground state.
The long half-life
value for the ground-state $\beta$ decay of $^{60}$Mn, 
in combination with the apparent direct feeding of the 
daughter $^{60}$Fe ground state, resulted in a log {\it ft}
value of 6.7, suggesting an isospin-forbidden 
Fermi decay of the $^{60}$Mn ground state.  Several 
examples of isospin-forbidden
$0^+ \rightarrow 0^+$ transitions are known \cite{ram1973}, 
all having log {\it ft} $> 6.5$.  Therefore, the $^{60}$Mn
ground-state $\beta$ decay was identified as a 
potential candidate for isospin-forbidden
$\beta$ decay.

Doubt regarding the long half-life of $^{60}$Mn$^{g}$
was reported by Schmidt-Ott {\it et al.} \cite{sch1993},
who set out to directly measure the multipolarity
of the IT in both $^{58,60}$Mn using conversion
electron spectroscopy.  At $A = 60$, the 
$M3$ multipolarity assignment to the 
272-keV $\gamma$ ray was 
confirmed based on the measured $\alpha _K$ value.
However, multiscaled $\beta$ singles measurements showed a 
long-lived $45.5 \pm 1.7$~s activity at $A = 60$ that was
associated with the decay of two known isomers in 
$^{120}$In \cite{che1978}.
The In activity was present as a doubly-ionized species 
from the ion source of the on-line mass separator and
appeared at $A = 60$ due to its having an identical 
$m/q$ ratio.
The initial half-life measurement of the $^{60}$Mn 
ground state by Bosch {\it et al.} \cite{bos1988} was 
also carried out using on-line mass separation and 
a possible contamination of the $^{60}$Mn $\beta$ spectrum
from $^{120}$In$^{2+}$ cannot be ruled out.  

We report a new measurement of the $\beta$-decay
half-life of the $^{60}$Mn ground state.  To
eliminate contributions from the $^{60}$Mn 
isomeric-state $\beta$ decay, we selectively populated the 
$^{60}$Mn ground state following the $\beta$ decay
of the even-even $^{60}$Cr ground state, with 
$J^{\pi} = 0^+$ and two independently
determined half-lives of $0.57 \pm 0.06$~s \cite{bos1988}
and $0.51 \pm 0.15$~s \cite{dor1996}.  

\section{Experimental Methods}
The parent $^{60}$Cr activity was produced following
projectile fragmentation of a 140-MeV/nucleon 
$^{86}$Kr beam at the National Superconducting Cyclotron 
Laboratory at Michigan State University.
The $^{86}$Kr primary beam, with an average 
beam current of 15~pnA, was 
incident onto a 376-mg/cm$^{2}$ 
thick Be target located at the object 
position of the A1900 fragment separator \cite{mor2003}.  
The secondary fragments 
of interest, including $^{60}$Cr, were selected in the 
A1900 separator using a 330~mg/cm$^2$ Al degrader
and 1\% momentum slits; both were located at the 
intermediate image of the device.  The dipole 
magnets of the A1900 
fragment separator were set to magnetic rigidities
$B \rho _1 = 4.239$~Tm and $B \rho _2 = 3.944$~Tm.
These same settings were used in the previously reported 
$\beta$-decay studies of $^{56}$Sc \cite{lid2004} and
$^{57}$Ti and $^{59}$V \cite{lid2005}.

The fully-stripped $^{60}$Cr fragments,
along with $^{56}$Sc, $^{57}$Ti, and $^{58,59}$V 
were implanted in a Double-Sided 
Si microstrip Detector (DSSD) with
thickness 1470~$\mu$m that is part 
of the NSCL $\beta$ counting system \cite{pri2003}.
Fragments were unambiguously identified by
a combination of multiple energy loss signals and 
time of flight.  A total of $2.75 \times 10^{5}$ 
$^{60}$Cr ions, composing 24\% of the secondary beam,
were implanted into the DSSD.

Fragment-$\beta$ correlations were established in software 
by requiring a high-energy implantation event in a single pixel 
of the DSSD, followed by a low-energy $\beta$ event
in the same or any of the eight nearest-neighbor pixels.
The differences between the absolute time stamps of 
correlated $\beta$ and implantation events were 
histogrammed to generate decay curves.
To suppress background, implantation 
events were rejected if they were 
not followed by a $\beta$ event within 5~s, or if 
they were followed by a second implantation within the 
same 5~s time interval.  The $\beta$-detection efficiency
was $\sim 30$\%.  Delayed $\gamma$ rays were measured with 
12 detectors from the MSU Segmented 
Germanium Array \cite{mue2001} 
arranged around the $\beta$ counting system.  The
$\gamma$-ray peak detection efficiency was 
5.3\% at 1~MeV.  The energy resolution for each 
of the Ge detectors 
was $\sim 3.5$~keV for 
the 1.3~MeV $\gamma$-ray
transition in $^{60}$Co.  Additional details regarding the 
$\beta$-delayed $\gamma$-ray techniques used with 
fast fragmentation beams at the NSCL are available in 
Ref.\ \cite{man2003}.

\section{Results and Discussion}
\begin{figure}[h]
\includegraphics{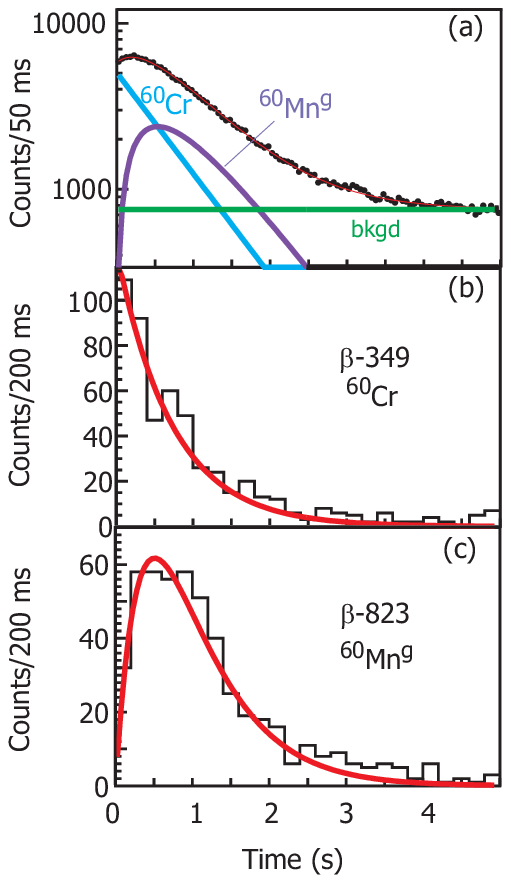}
\caption{(Color online) (a) Decay curve for $^{60}$Cr from 
fragment-$\beta$ correlations.  The data were 
fitted with an exponential parent decay, an exponential
daughter growth and decay, and a linear background. The 
decay curves for fragment-$\beta \gamma$ correlations 
with gates on $\gamma$-ray transitions with energies
349 and 823 keV are shown in (b) and (c), respectively.}
\label{fig1:cr60decaycurve}
\end{figure}

The decay curve for $\beta$ events that occurred within 
5~s of a $^{60}$Cr implantation event is shown 
in Fig.\ \ref{fig1:cr60decaycurve}(a).  From the shape of the 
decay curve
below 100~ms, it is apparent that the daughter $^{60}$Mn 
has a shorter half-life than $^{60}$Cr.  The 
decay curve of Fig.\ \ref{fig1:cr60decaycurve}(a) was fitted 
with a function that considered the exponential 
decay of the $^{60}$Cr parent, the exponential 
growth and decay of the $^{60}$Mn daughter, 
and a linear background.  A half-life value of $0.49 \pm 0.01$~s 
was deduced for the $^{60}$Cr parent, in reasonable agreement 
with previous measurements by Bosch {\it et al.} \cite{bos1988}   
$0.57 \pm 0.06$~s and D\"{o}rfler {\it et al.} \cite{dor1996}
$0.51 \pm 0.15$~s.  A half-life value of $0.28 \pm 0.02$~s was 
deduced for the $^{60}$Mn ground-state, 
in stark contrast to the
long half-life of $51 \pm 6$~s deduced by Bosch 
{\it et al.} \cite{bos1988}.

\begin{figure}[h]
\includegraphics{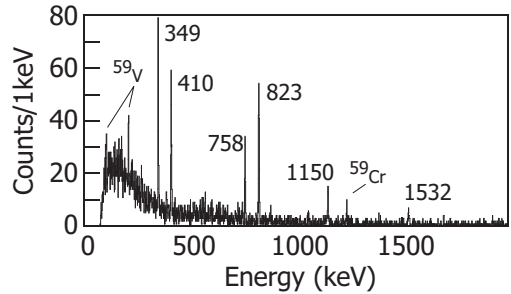}
\caption{Delayed $\gamma$-ray spectrum for $\beta$-decay
events that occurred within 1~s of a 
$^{60}$Cr implantation event.  Peaks are marked 
by their transition energy in keV.  Contaminant lines 
from the $A =59$ decay chain were also observed due
to some overlap in the particle identification spectrum.}
\label{fig2:cr60delayedgammas}
\end{figure}

The $\beta$-delayed $\gamma$-ray spectrum for 
events that occurred within
1~s of a $^{60}$Cr implantation event 
within the DSSD is given
in Fig.\ \ref{fig2:cr60delayedgammas}.  $\gamma$ rays 
evident in this spectrum correspond to short-lived
$\beta$-decay events.  Since the ground states of 
both $^{60}$Cr and $^{60}$Mn were found to have half-life values 
of less than 1~s, their corresponding delayed $\gamma$ rays 
should be present in this spectrum.  The assignment 
of $\gamma$-ray transitions to either the parent or 
daughter decay was accomplished by analysis of the 
fragment-$\beta \gamma$ decay curves.  The 
823-keV $\gamma$ ray had been previously assigned 
to the deexcitation of the first excited $2^+$ state in 
$^{60}$Fe. The fragment-$\beta \gamma$ 
decay curve for this transition, 
illustrated in Fig.\ \ref{fig1:cr60decaycurve}(c),
exhibits the growth of $^{60}$Mn$^g$ (from the $^{60}$Cr parent decay) 
as well as its presently-determined $T_{1/2} = 0.28$~s decay.
The decay curves gated on
the delayed $\gamma$ rays with energies 
1150 and 1532 keV showed similar structure and have also
been assigned in the present work to the deexcitation 
of levels in $^{60}$Fe.
The other three delayed $\gamma$-ray transitions 
in Fig.\ \ref{fig2:cr60delayedgammas} all reveal
half-lives consistent with the decay of the $^{60}$Cr 
parent.  As an example, the $\gamma$-ray gated decay curve 
derived for the 349-keV transition is 
presented in Fig.\ \ref{fig1:cr60decaycurve}(b). The three
transitions with energies 349, 410, and 758 keV have, in
the present work, been identified as depopulating 
excited states in $^{60}$Mn.

The new half-life value of $0.28 \pm 0.02$~s for the 
decay of $^{60}$Mn$^g$ is not consistent with the 
$0^+$ spin and parity quantum numbers previously assigned to the 
$^{60}$Mn ground state.  Such a short half-life and 
direct $\beta$-decay feeding to the $^{60}$Fe ground state would 
purport a large $B$(Fermi) value that is unexpected for 
an isospin-forbidden $0^+ \rightarrow 0^+$ Fermi transition.    
A review of the experimental basis for the spin and parity 
assignments to the ground and isomeric states in $^{60}$Mn,
therefore, is warranted.  The $\Delta J = 3$ spin difference between 
the ground and isomeric states is firmly established 
by the conversion electron measurements of Schmidt-Ott
{\it et al.} \cite{sch1993}.  However, the
initial $3^+$ spin and parity assignment to the isomeric state
can be called into question, since this assignment is 
based solely on the apparent log $ft$ values indicating allowed
$\beta$ decays to the $2^+_2$, $3^+_1$ and $4^+_1$ states in 
$^{60}$Fe \cite{nor1978}.  The absolute $\beta$ intensity
to the $2^+_2$ level in $^{60}$Fe was reported to be
$(6.7 \pm 2.0)$\%, but with a large $Q_{\beta }$ value window
of 8.2~MeV \cite{aud2003} this apparent $\beta$ feeding
should only be considered an upper limit since the presence
of unobserved transitions from higher-energy states in 
$^{60}$Fe cannot be ruled out \cite{har1977}.        
An assignment of $J^{\pi} = 4^+$, therefore, cannot be excluded
for the $^{60}$Mn isomeric state.  The adoption of 
$J^{\pi} = 4^+$ for $^{60}$Mn$^m$ would also change the 
proposed ground state spin and parity of $^{60}$Mn to 
$J^{\pi} = 1^+$ due to the known $M3$ multipolarity for the 
272-keV IT between the two states.  The apparent direct $\beta$-decay feeding 
to the ground and two excited $0^+$ states in 
$^{60}$Fe (see below) would be consistent with 
allowed Gamow-Teller decay if the parent state had $J^{\pi} = 1^+$.

\begin{figure}[h]
\includegraphics{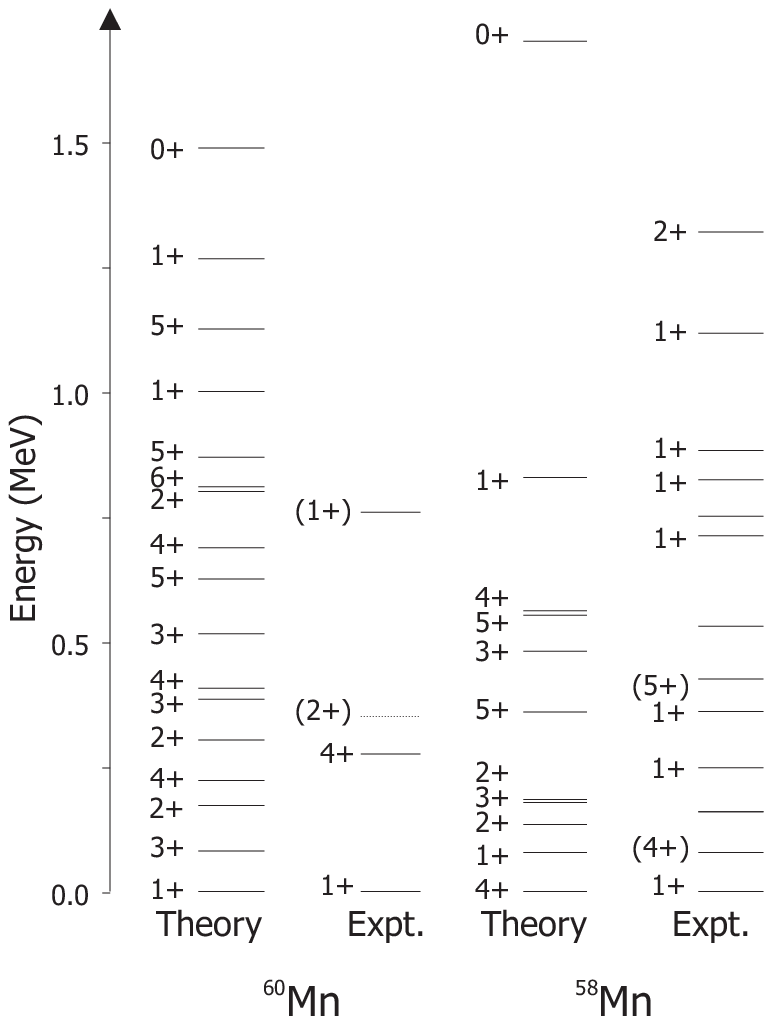}
\caption{Comparison of the known low-energy
level structure of $^{58}$Mn and $^{60}$Mn with the
results of shell model calculations in the
full $pf$-shell model space using the 
GXPF1 interaction.  Only the shell model results
for the first three (two) states for each spin and 
parity up to $J^{\pi} = 5^+$ for $^{60}$Mn ($^{58}$Mn) 
are shown.  Experimental results for $^{58}$Mn are taken
from Ref.\ \cite{bha1997}, while the data for 
$^{60}$Mn are from the present work.}
\label{fig5:mn60theory-full}
\end{figure}

To further support the proposed changes in the 
spin and parity assignments of $1^+$ and $4^+$
to the ground and 
isomeric states of $^{60}$Mn, respectively, the low-energy 
levels of this nuclide were calculated in the
full $pf$-shell model space with the 
GXPF1 interaction \cite{hon2002,hon2004}.  This interaction   
is derived from a microscopic calculation based
on renormalized G matrix theory with the 
Bonn-C interaction and refined by a systematic
adjustment of the important linear combinations of
two-body matrix elements to low-lying states in nuclei
from $A=47$ to $A=66$.  The 
GXPF1 interaction has met with success in describing
the $\beta$-decay properties and low-energy levels of 
the neutron-rich $\pi f_{7/2} - \nu pf$ shell nuclei 
\cite{man2003,lid2004,lid2005,jan2002}.  
The {\textsc OXBASH} \cite{oxbash} 
and {\textsc CMICHSM} \cite{hor2003} codes
were used to generate the results 
for $^{60}$Mn shown in Fig.\ \ref{fig5:mn60theory-full}.
The ground state is predicted to have $J^{\pi} = 1^+$,
and the first $4^+$ level is calculated at 224~keV.  The first 
$J^{\pi} = 0^+$ level is expected at an energy
more than 1.4~MeV above the ground state.  
The wavefunctions of the lowest-energy $1^+$ and $4^+$ states 
are predominately associated with the 
$\pi (f_{7/2}) - \nu (f_{5/2})$ multiplet.  In retrospect,
the coupling scheme which would produce a $0^+$ state from the
shell model orbitals expected to be occupied by 
the valence protons and neutrons in $^{60}$Mn is not 
obvious; the $\pi (f_{7/2})$ orbital is half-filled
and valence neutrons should occupy the $p_{1/2}$
and $f_{5/2}$ orbitals so that $j_p - j_n = 0$ 
is not possible \cite{nor1951,bre1960}.

Shell model calculations were also completed in a 
similar manner
for $^{58}$Mn.  Bosch {\it et al.} \cite{bos1988} initially
proposed $J^{\pi} = 0^+$ for the ground state of $^{58}$Mn.  
However, the small comparative half-life (apparent log {\it ft}
value of 4.9) for direct decay to the $0^+$ ground state 
of the $^{58}$Cr was not suggestive of isospin-forbidden
$\beta$ decay.  The results of the shell model 
calculations in the full $pf$-shell model space 
suggest that the first $0^+$ state should reside 
more than 1.5~MeV above the ground state (see Fig.\ 
\ref{fig5:mn60theory-full}.)  Indeed, 
spin and parity of $1^+$ have been adopted
for the $^{58}$Mn ground state in the most 
recent data compilation \cite{bha1997}. 
The calculated ground
and first excited states, with $J^{\pi} = 4^+$ and
$1^+$, respectively, are separated in energy
by 75~keV.  Experimentally, these states are observed
in reverse order and with a similar energy separation
of 72 keV.  The order of states separated by such 
a small energy difference cannot be accurately predicted
by the GXPF1 interaction, which has an $rms$ error of 168~keV 
\cite{hon2004}. However, the calculated $\beta$ decay half-life
of 2.75~s for the lowest $1^+$ level in $^{58}$Mn is in
good agreement with the experimental half-life value
of $3.0 \pm 0.1$~s \cite{war1969} of $^{58}$Mn$^{g}$,
suggesting that the adopted order of states, with $1^+$ as the 
ground state and $4^+$ as the first excited state, is correct.
 
\begin{figure}[h]
\includegraphics{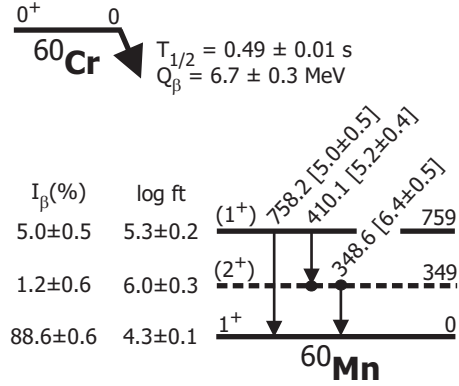}
\caption{Partial level scheme for $^{60}$Mn 
populated following the $\beta$ decay 
of $^{60}$Cr.  The number in brackets following a 
$\gamma$-ray transition energy is the 
absolute $\gamma$-ray intensity.  The $Q$ value was
deduced from data in Ref.\ \cite{aud2003}.  
Observed coincidences are represented as filled 
circles.  Absolute $\beta$-decay intensities and apparent
log {\it ft} values to each state in $^{60}$Mn are 
given on the left-hand side of the figure.  The 349-keV level
is dashed only because of the uncertainty in the
ordering of the 348.6- and 410.1-keV $\gamma$ rays.
This level could instead be at 410 keV.}
\label{fig3:cr60levelscheme}
\end{figure}

\begin{figure}[h]
\includegraphics{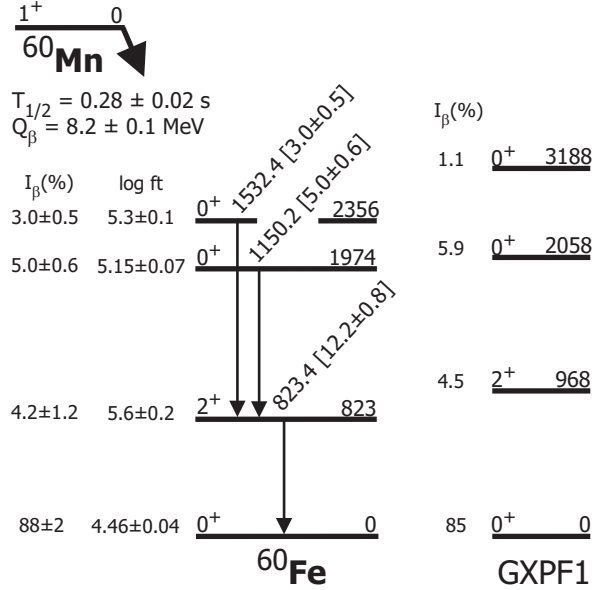}
\caption{Partial level scheme for $^{60}$Fe 
populated following the $\beta$ decay 
of $^{60}$Mn$^g$.  The number in brackets following a 
$\gamma$-ray transition energy is the
relative $\gamma$-ray intensity.  The $Q$ value was
deduced from data in Ref.\ \cite{aud2003}.  Apparent 
$\beta$ feedings and log {\it ft} values are given
to the left of the proposed level scheme.  The 
results of shell model calculations using the 
GXPF1 interaction are presented to the right of the 
experimental level scheme.}
\label{fig4:fe60levelscheme}
\end{figure}

Further details regarding the 
$\beta$-decay properties of both 
$^{60}$Cr and $^{60}$Mn$^{g}$ are given in 
Fig.\ \ref{fig3:cr60levelscheme}
and Fig.\ \ref{fig4:fe60levelscheme}, respectively.
For $^{60}$Cr, a new half-life value of 
$0.49 \pm 0.01$~s was deduced from the decay
curves gated on the $\gamma$-ray transitions 
with energies of 349, 410, and 758 keV.  This value 
agrees with previously-reported half-life values of 
$0.57 \pm 0.06$~s \cite{bos1988} 
and $0.51 \pm 0.15$~s \cite{dor1996}.
The 349- and 410-keV $\gamma$ rays are coincident; 
however, the ordering
of the two transitions in the level scheme of 
$^{60}$Mn cannot be resolved by the present
data set.  The 758-keV 
$\gamma$ ray is the sum energy of the 
349- and 410-keV transitions and has been
placed as directly feeding the $^{60}$Mn ground
state.  The majority of the $\beta$ decay
of $^{60}$Cr directly populates the 
$^{60}$Mn ground state.  Apparent $\beta$ branching
to the two new excited levels in $^{60}$Mn is 
also observed, but the feeding intensities 
quoted in Fig.\ \ref{fig3:cr60levelscheme} 
should be considered as upper limits, 
since the $Q_{\beta}$ window is large and 
possible unobserved transitions depopulating
into these excited states could reduce the 
apparent feedings.  Tentative spin and parity 
values of $1^+$ and $2^+$ have been assigned to the 
759-keV and 349-keV levels, respectively, based on the 
$\beta$ feeding patterns and the correspondence
with the shell model results presented in
Fig.\ \ref{fig5:mn60theory-full} and
Table  \ref{tab1:theorydecay}.

Previous studies of the $\beta$-decay of 
$^{60}$Mn$^{g}$ suggested that no excited 
states were populated with any significant 
$\beta$ intensity \cite{bos1988}.  Three delayed
$\gamma$ rays have been assigned in this work
to the decay of $^{60}$Mn$^{g}$ with energies
823, 1150, and 1532 keV.  The $\gamma$-ray 
line at 823 keV had previously been 
assigned to the $2^+_1 \rightarrow 0^+_1$ 
transition in $^{60}$Fe.  The other two $\gamma$ rays
with energies 1150 and 1532 keV are proposed to 
depopulate the $0^+_2$ and $0^+_3$ states, respectively.
The two excited $0^+$ states were previously known
from transfer reaction work, and the $J = 0$ assignment
is based on $\ell =0$ two-neutron transfer to
the $^{58}$Fe target \cite{wat1986}.   A 1150-keV
$\gamma$ ray was similarly identified as deexciting
the 1974-keV $0^+_2$ level based on a study of the
$^{58}$Fe(t,p$\gamma$) reaction by Warburton {\it et al.} 
\cite{war1977}.  While Runte {\it et al.} \cite{run1985}
observed the 1150-keV $\gamma$ ray when studying the 
$\beta$ decay of the now-adopted $4^+$ isomeric
state in $^{60}$Mn, it is plausible that 
they unknowingly monitored a mixture of both the 
high- and low-spin $\beta$-decaying states.   

The branching ratios
for both the $^{60}$Cr and 
$^{60}$Mn$^{g}$ $\beta$ decay 
were also calculated in the full $pf$-shell model
space with the GXPF1 interaction and the 
CMICHSM code \cite{hor2003}. There is excellent agreement 
between the experimental 
observations and the shell model results in both 
instances, as seen from 
Table \ref{tab1:theorydecay}.  The 
predicted half-lives take into consideration
a quenching factor of 0.7 for 
the Gamow-Teller strength in this region \cite{pov2001}.
Ground state to ground state $\beta$ decay 
dominates the decay pathway for both parent 
and daughter nuclides, which corroborates
the experimental observations.      

\section{Summary}
A new half-life value of $0.28 \pm 0.02$~s 
has been deduced for the ground state of 
$^{60}$Mn, which is significantly shorter than
the previously adopted value of $51 \pm 6$~s.
The prospect of isospin-forbidden 
$0^+ \rightarrow 0^+$ Fermi $\beta$ decay for the 
$^{60}$Mn ground state has been ruled out by the 
new, short half-life value.  New spin and parity
assignments of $1^+$ and $4^+$ have been 
adopted for the ground and the isomeric $\beta$-decaying
states in $^{60}$Mn, respectively, supported by the 
deduced $\beta$-decay properties reported here
and by the results of shell model calculations 
using the GXPF1 interaction in the full
$pf$-shell model space.  Two new low-energy 
levels have been identified in $^{60}$Mn
and evidence for direct $\beta$ feeding of 
excited $0^+$ states in $^{60}$Fe is reported 
for the first time.   

\begin{acknowledgments}
The work was supported in
part by the National Science Foundation Grants
PHY-01-10253, PHY-97-24299, PHY-01-39950, and
PHY-02-44453, and by the US Department of 
Energy, Office of Nuclear Physics, under 
contract W-31-109-ENG-38.  
The authors would like to thank the NSCL operations
staff for providing the primary and
secondary beams for this experiment.  The 
authors also thank the members
of the NSCL $\gamma$ group who helped
with the set up of the Segmented Germanium Array.
The DSSD used for these measurements was 
provided by K.~Rykaczewski (ORNL).  The shell
model calculations were performed on the 
computer systems provided by the 
High Performance Computing Center
at Michigan State University.
 
\end{acknowledgments}

\begin{table}[h]
\caption{Calculated $\beta$-decay branching ratios
for the ground-state decays of $^{60}$Cr and 
$^{60}$Mn.  Details regarding the shell model calculations in
the full $pf$ shell with the 
GXPF1 interaction are given in the text.}
\label{tab1:theorydecay}
\begin{ruledtabular}
\begin{tabular}{cccccccc}
\multicolumn{4}{c}{$^{60}$Cr (0$^+$) $\rightarrow$ $^{60}$Mn} &
\multicolumn{4}{c}{$^{60}$Mn (1$^+$) $\rightarrow$ $^{60}$Fe} \\
\cline{1-4} \cline{5-8}
$E_f$(keV)  &$J^{\pi}_f$     &$I^{expt.}_{\beta}$(\%)
& $I^{theory}_{\beta}$(\%)    &
       $E_f$(keV)  &$J^{\pi}_f$     
&$I^{expt.}_{\beta}$(\%)& $I^{theory}_{\beta}$(\%)     \\
\hline
0     &     $1^+$   &$88.6\pm 0.6$           &91.3              &
        0      &$0^+$         &$88 \pm 2$           &85              \\
(349) &    ($2^+$)  &$1.2\pm0.6$             &                  &
        823    &$2^+$         &$4.2\pm 1.2$         &4.5             \\
758   &    ($1^+$)  &$5.0 \pm 0.5$           &3.6               &
       1974    &$0^+$         &$5.0\pm 0.6$         &5.9             \\
      &             &                        &                  &
       2356    &$0^+$         &$3.0\pm 0.5$         &1.1             \\
Sum   &             &                        &94.9              &
       Sum     &              &                     &96.5            \\
\hline
T$_{1/2}^{expt.}$ & \multicolumn{3}{c}{$0.49 \pm 0.01$ s} &
\multicolumn{4}{c}{$0.28 \pm 0.02$ s} \\
T$_{1/2}^{theory}$ & \multicolumn{3}{c}{0.25 s} &
\multicolumn{4}{c}{0.21 s} \\
\end{tabular}
\end{ruledtabular}
\end{table}

\begin{thebibliography}{99}

\bibitem{nor1978} E.B.~Norman, C.N.~Davids, M.J.~Murphy, 
and R.C.~Pardo,
Phys.\ Rev.\ C {\bf 17}, 2176 (1978).

\bibitem{run1985} E.~Runte, K.-L.~Gippert, W.-D.~Schmidt-Ott,
P.~Tidemand-Petersson, L.~Ziegeler, R.~Kirchner, O.~Klepper,
P.O.~Larsson, E.~Roeckl, D.~Schardt, N.~Kaffrell, P.~Peuser,
M.~Bernas, P.~Dessagne, M.~Langevin, and K.~Rykaczewski,
Nucl.\ Phys.\ {\bf A441}, 237 (1985).

\bibitem{bos1988} U.~Bosch, W.-D.~Schmidt-Ott, 
E.~Runte, P.~Tidemand-Petersson, P.~Koschel, 
F.~Meissner, R.~Kirchner, O.~Klepper, E.~Roeckl, K.~Rykaczewski, 
and D.~Schardt, Nucl.\ Phys.\ {\bf A477}, 89 
(1988).

\bibitem{ram1973} S.~Raman and N.B.~Gove, 
Phys.\ Rev.\ C {\bf 7}, 1995 (1973).

\bibitem{sch1993} W.-D.~Schmidt-Ott, K.~Becker, U.~Bosch-Wicke,
T.~Hild, F.~Meissner, R.~Kirchner, E.~Roeckl, and K.~Rykaczewski,
Proc.\ 6th Intern.\ Conf.\ on Nuclei Far from Stability and 
9th Intern.\ Conf.\ on Atomic Masses and Fundamental Constants, 
Bernkastel-Kues, Germany, R.Neugart, A.Wohr, Eds.\, p.627 (1993).

\bibitem{che1978} H.C.~Cheung, H.~Huang, B.N.~Subba Rao, 
L.~Lessard, and J.K.P.~Lee, J.\ Phys.\ G {\bf 4}, 1501 (1978).

\bibitem{dor1996} T.~D\"{o}rfler, W.-D.~Schmidt-Ott, T.~Hild, 
T.~Mehren, W.~B\"{o}hmer, P.~M\"{o}ller, B.~Pfeiffer,
T.~Rauscher, K.-L.~Kratz, O.~Sorlin, V.~Borrel, S.~Gr\'{e}vy,
D.~Guillemaud-Mueller, A.C.~Mueller, F.~Pougheon, 
R.~Anne, M.~Lewitowicz, A.~Ostrowsky, M.~Robinson, and
M.G.~Saint-Laurent,
Phys.\ Rev.\ C {\bf 54}, 2894 (1996).

\bibitem{mor2003} D.J.~Morrissey, B.M.~Sherrill, M.~Steiner,
A.~Stolz, and I.~Wiedenh\"{o}ver,  Nucl.\ Instrum.\ Methods 
Phys.\ Res.\ B {\bf 204}, 90 (2003).

\bibitem{lid2004} S.N.~Liddick , P.F.~Mantica, R.~Broda, B.A.~Brown, 
M.P.~Carpenter, A.D.~Davies, B.~Fornal, T.~Glasmacher, D.E.~Groh, M.~Honma, 
M.~Horoi, R.V.F.~Janssens, T.~Mizusaki, D.J.~Morrissey, A.C.~Morton, W.F.~Mueller, 
T.~Otsuka, J.~Pavan, H.~Schatz, A.~Stolz, S.L.~Tabor, B.E.~Tomlin, and M.~Wiedeking,
Phys.\ Rev.\ C {\bf 70}, 064303 (2004). 

\bibitem{lid2005} S.N.~Liddick, P.F.~Mantica, R.~Broda, B.A.~Brown, M.P.~Carpenter, 
A.D.~Davies, B.~Fornal, M.~Horoi, R.V.F.~Janssens, A.C.~Morton, W.F.~Mueller, 
J.~Pavan, H.~Schatz, A.~Stolz, S.L.~Tabor, B.E.~Tomlin, and M.~Wiedeking
Phys.\ Rev.\ C {\bf 72}, 054321 (2005). 

\bibitem{pri2003} J.I.~Prisciandaro, A.C.~Morton, and P.F.~Mantica,
Nucl.\ Instrum.\ Methods Phys.\ Res.\ A {\bf 505}, 140 (2003).

\bibitem{mue2001} W.F.~Mueller, J.A.~Church, T.~Glasmacher, D.~Gutknecht, 
G.~Hackman, P.G.~Hansen, Z.~Hu, K.L.~Miller, P.~Quirin, 
Nucl.\ Instrum.\ Methods Phys.\ Res.\ A {\bf 466}, 492 (2001).

\bibitem{man2003} P.F.~Mantica, A.C.~Morton, B.A.~Brown, 
A.D.~Davies, T.~Glasmacher, D.E.~Groh, S.N.~Liddick,
D.J.~Morrissey, W.F.~Mueller, H.~Schatz, A.~Stolz, 
S.L.~Tabor, M.~Honma, M.~Horoi, and T.~Otsuka, Phys.\ Rev.\ C {\bf 67},
014311 (2003).

\bibitem{aud2003} G.~Audi, A.H.~Wapstra, and 
C.~Thibault, Nucl.\ Phys.\ {\bf A729}, 337 (2003).

\bibitem{har1977} J.C.~Hardy, L.C.~Carraz, B.~Jonson, 
and P.G.~Hansen, Phys.\ Lett.\ {\bf 71B}, 307 (1977).

\bibitem{hon2002} M.~Honma, T.~Otsuka, B.A.~Brown, and T.~Mizusaki,
Phys.\ Rev.\ C {\bf 65}, 061301(R) (2002).

\bibitem{hon2004} M.~Honma, T.~Otsuka, B.A.~Brown, and T.~Mizusaki,
Phys.\ Rev.\ C {\bf 69}, 034335 (2004).

\bibitem{jan2002} R.V.F.~Janssens, B.~Fornal, P.F.~Mantica,
B.A.~Brown, R.~Broda, P.~Bhattacharyya, M.P.~Carpenter, M.~Cinausero, 
P.J.~Daly, A.D.~Davies, T.~Glasmacher, Z.W.~Grabowski, D.E.~Groh, 
M.~Honma, F.G.~Kondev, W.~Krolas, T.~Lauritsen, S.N.~Liddick, S.~Lunardi, 
N.~Marginean, T.~Mizusaki, D.J.~Morrissey, A.C.~Morton, W.F.~Mueller, 
T.~Otsuka, T.~Pawlat, D.~Seweryniak, H.~Schatz, A.~Stolz, S.L.~Tabor, 
C.A.~Ur, G.~Viesti, I.~Wiedenhoever, and J.~Wrzesinski, 
Phys.\ Lett.\ B {\bf 546}, 55 (2002).

\bibitem{oxbash} B.A.~Brown, A.~Etchegoyen and W.D.M.~Rae, 
The computer code OXBASH, MSU-NSCL Report No.\ 524, 1998.

\bibitem{hor2003} M.~Horoi, B.A.~Brown, and V.~Zelevinsky,
Phys.\ Rev.\ C {\bf 67}, 034303 (2003).

\bibitem{nor1951} L.~Nordheim, Rev.\ Mod.\ Phys.\ {\bf 23}, 322 (1951).

\bibitem{bre1960} M.H.~Brennan and A.M.~Bernstein, Phys.\
Rev.\ {\bf 120}, 927 (1960).

\bibitem{bha1997} M.R.~Bhat, Nucl.\ Data Sheets {\bf 80}, 789 (1997).

\bibitem{war1969} T.E.~Ward, P.H.~Pile, and P.K.~Kuroda, 
Phys.\ Rev.\ {\bf 182}, 1186 (1969).

\bibitem{wat1986} D.L.~Watson and H.T.~Fortune, Nucl.\ Phys.\
{\bf A448}, 221 (1986). 

\bibitem{war1977} E.K.~Warburton, J.W.~Olness, 
A.M.~Nathan, J.J.~Kolata, and J.B.~McGrory,
Phys.\ Rev.\ C {\bf 16}, 1027 (1977).

\bibitem{pov2001} A.~Poves, J.~Sanchez-Solano, 
E.~Caurier, and F.~Nowacki, Nucl.\ Phys.\
{\bf A694}, 157 (2001).
 

\end{thebibliography}
\end{document}